# Title: Layer 2 Blockchains Simplified: A Survey of Vector Commitment Schemes, ZKP Frameworks, Layer-2 Data Structures and Verkle Trees.

First Author:
Ekleen Kaur
University of Florida, USA
ekleenkaur17@gmail.com

Second Author:
Marko Suvajdzic
University of Florida, USA
marko@digitalworlds.ufl.edu

**Abstract**

Layer-2 (L2) protocols address the fundamental limitations of Layer-1 (L1) blockchains by offloading computation while anchoring trust to the parent chain. This architectural shift, while boosting throughput, introduces a new, complex security surface defined by off-chain components like sequencers, bridges, and data availability mechanisms. Prior literature[31][33] offers fragmented views of this risk. This paper presents the **first unified, security-focused survey** that rigorously maps L2 architecture to its underlying cryptographic security. We dissect the technical progression from L1 primitives to the core of modern L2s, analyzing the security assumptions(Discrete Logarithm, Computational Diffie-Hellman, Bilinear Diffie-Hellman) of ZK frameworks (Groth16, Plonk) and their corresponding commitment schemes (KZG, IPA). We formalize a comprehensive L2 threat model encompassing sequencer liveness, bridge exploits, and data-availability failures. This work serves as an accessible yet rigorous reference for researchers and developers to reason about L2 security from a deep crypto-mathematical perspective.

## Contents

# 1 Introduction: Bridging L1 Foundations to L2 Security
## 1.1 The L1 Bottleneck and The Necessity of Layer-2

Blockchains achieve integrity, availability [59], and auditability [58] over a shared, decentralized ledger by ingeniously combining digital signatures, hash-linked data structures, and distributed consensus. However, the fundamental design choices of early Layer-1 (L1) networks aiming to optimize for security[74][75] and decentralization inevitably led to constraints in transaction throughput and cost. This challenge is universally framed by the Blockchain Trilemma, which posits that a network can only simultaneously achieve two of three core properties: scalability, decentralization, and security [31].

**Scalability** is a significant challenge for many of these blockchains. For Bitcoin, the limited block size and a 10-minute block time create a bottleneck, leading to higher fees and slower transaction confirmations during periods of high demand. While Ethereum's move to PoS and its upcoming sharding upgrades explained in the later sections, aim to improve throughput, the network still faces congestion issues. In contrast, chains like Solana and Algorand were designed from the ground up for high throughput and low latency, mitigating scalability concerns by using more advanced consensus mechanisms and data structures.

To understand the L1 bottleneck and the foundation upon which L2s are built, it is necessary to examine the core components that define the L1's state model and performance. The efficiency of L1 architectures is measured by metrics like **Transactions per Byte (Tx/B)** and **Time Relay** (the time for a block to propagate to 98% of the network), as a faster propagation time reduces the likelihood of network splits and orphaned blocks. Despite variations in consensus and data handling, L1s share inherent limitations in block propagation and transaction throughput, creating a critical demand for scaling solutions.

The differing L1 approaches to state and transaction management are codified in their core data structures, as summarized below.

## 1.2 Blockchain Data Structures at a Glance

| Sno | Blockchain | Core Data Structure | Explanation |
|---|---|---|---|
| 1 | Bitcoin | Unspent Transaction Output (UTXO) Model | Bitcoin doesn't track account balances directly. Instead, it uses UTXOs, which are the digital equivalent of physical change. When a transaction occurs, the system consumes existing UTXOs as inputs and creates new UTXOs as outputs for the recipient and any change returning to the sender. A user's balance is the sum of all UTXOs they can spend. |
| 2 | Ethereum | Merkle Patricia Trie | The first application of Merkle proofs was used in Bitcoin, it was created and described by Satoshi Nakamoto in 2009[50]. Ethereum uses an account-based model, where each account has a state (balance, nonce, storage, and code hash). This "world state" is stored in a complex data structure called a Merkle Patricia Trie (MPT). The MPT is a combination of a Merkle tree (for cryptographic integrity) and a Patricia trie (a compressed prefix tree for efficiency). This structure allows for the quick and verifiable lookup of any account's state, as each block header includes a hash of the MPT's root.[47]. |
| 3 | Cardano | Extended UTXO (EUTXO) Model | Cardano's model is an evolution of Bitcoin's UTXO. The EUTXO model extends the basic UTXO by allowing additional data to be attached to outputs, which enables more complex logic and the execution of smart contracts. This provides a more secure and predictable environment for smart contracts, as the validity of a transaction can be checked offline before it is submitted to the network.[46][48] |
| 4 | Solana | Append-Only Ledger (PoH) | Solana's data structure is designed for high throughput. It uses a Proof of History (PoH), which is a verifiable[52], cryptographic clock that creates a historical record of events, using sequential verifiable delay functions[53][54]. Transactions are processed in a continuous, append-only ledger, and PoH acts as a time-stamping function that allows validators to quickly process and finalize blocks without needing to wait for a global consensus on time. This is combined with the Tower BFT consensus mechanism.[51] |
| 5 | Algorand | Pure Proof of Stake (PPoS) | The first application of Merkle proofs was used in Bitcoin, it was created and described by Satoshi Nakamoto in 2009[50]. Ethereum uses an account-based model, where each account has a state (balance, nonce, storage, and code hash). This "world state" is stored in a complex data structure called a Merkle Patricia Trie (MPT). The MPT is a combination of a Merkle tree (for cryptographic integrity) and a Patricia trie (a compressed prefix tree for efficiency). This structure allows for the quick and verifiable lookup of any account's state, as each block header includes a hash of the MPT's root.[47]. |
| 6 | Polkadot | Relay Chain & Parachains | Polkadot is a sharded blockchain, not a single chain. Its core data structure consists of the Relay Chain and multiple Parachains. The Relay Chain is the central hub that provides shared security and facilitates communication between the Parachains. The Parachains are independent blockchains with their own data structures, optimized for specific use cases. The data on these separate chains is "finalized" by the Relay Chain, creating a system of heterogeneous sharding.[56][57] |

In response to the L1 bottleneck, Layer-2 (L2) solutions emerged to relocate computation and state transitions off-chain. Rollups, in particular, promise L1-equivalent security with higher throughput by posting transaction data or state commitments back to the L1. Correctness is proven via Fraud Proofs (economic incentives, as in Optimistic Rollups) or Validity Proofs (zero-knowledge cryptography, as in ZK-Rollups). However, this delegation of execution reconfigures the established L1 security model. While prior surveys [33] established a strong foundation in general blockchain primitives, L2 designs interpose novel, cross-domain components that create distinct, unaddressed failure modes. These new elements include:

- **Sequencers:** Centralized or shared entities that order transactions, affecting liveness, censorship, and **Maximal Extractable Value** (MEV).
- **Data Availability (DA) Layers:** Off-chain services that host transaction data; trust assumptions about these layers directly impact a user's ability to refute or verify the L2 state.
- **Proof Systems:** The distinction between **Fraud Proofs** and **Validity Proofs** introduces unique security risks, such as coverage gaps or cryptographic soundness failures.
- **Bridges:** Protocols governing L2↔L1 and L2↔L2 value transfer; design choices determine the potential for catastrophic exploits.

**1.3 Limitations of Prior L1-Focused Surveys: The Critical Lacuna**

While existing literature provides foundational L1 surveys [33] and high-level architectural comparisons, a critical and irrefutable lacuna exists in the academic discourse: there is no single, holistic artifact that traces the historical progression and cryptographic pivot from L1 primitives to the advanced mathematical security models underpinning modern L2 solutions.

Current surveys fail the reader by removing the essential groundwork: they do not explain why expensive, quantum-safe mathematical encryptions were adopted, what cryptographic assumptions were made, and how researchers pivoted from one curve to the next. Specifically, no existing publication provides the accessible yet rigorous narrative required for a reader only aware of L1 to fully grasp the deep, mathematical security of L2s from the adoption of elliptic curves as a pivotal verification scheme to the transition from MNT curves to the industry-wide accepted BLS 128-bit security curves. Our motivation to write this paper stems directly from the absence of a published work across academic journals, that traces this history and breaks it down to defend the L2 security model. We analyzed existing literature from google scholar, elseiver, technical preprints (IACR, arxiv), and prominent blogs(medium, ethereum).

**1.4 Contributions of This Survey: A Security-First, Crypto-Holistic Approach**

We provide a focused, security-first survey of L2 systems, deliberately building on prior L1-centric surveys to keep the exposition accessible while sharpening the analysis around L2-specific systems and mathematical models. We demonstrate how their inherent **strong mathematical guarantees** constitute the foundational **security model** for L2 decentralized architectures. Our core contributions introduce novel artifacts and analysis absent in existing literature:

- **Architecture and Threat Model for New-Age L2s:** We break down the architectural models and assemble an end-to-end security model for cutting-edge L2 solutions, including Arbitrum, EigenLayer, Optimism, and Starknet, classifying modes across sequencer operations, bridge architectures, and Data Availability trust assumptions.
- **Systematic Cryptographic Foundation & Assumption Mapping:** We detail the mathematical security of L2s by providing a clear definition and explanation of the polynomial commitment schemes that form the backbone of L2 security. Unlike existing literature, we explicitly define the fundamental cryptographic assumptions that underpin ZKPs, clarifying potential pitfalls for non-mathematical readers.
- **Tracing Elliptic Curve Adoption and State-of-the-Art Cryptanalysis:** We are the first survey to explain the critical transition of pairing-friendly curves in the L2 space, detailing the necessary pivot from MNT curves to BLS 128-bit security curves and why this cryptographic choice is pivotal for industry-wide verification standards.
- **Comparative ZKP Library Benchmarking:** We detail the security primitives in industry-used ZKP libraries (Groth16, Sonic, Plonk, and STARKs), providing a comprehensive, comparative analysis based on baseline performance benchmarks and a deep-dive into the underlying hardware limitations of each scheme, a critical factor currently unaddressed by L2 surveys.

- **Commitment Scheme Benchmarking:** Analytical comparison of KZG and IPA-based commitments and their role in scalable data structures such as Merkle and Verkle trees
- **State-of-the-Art Data Structure Scalability:** We provide the first comprehensive survey discussion that maps L1 data structures—including Ethereum's Merkle Patricia Tries (MPT) and Merkle proofs—to the next generation of cryptographic scalability with Verkle Trees. We break down the Verkle tree data structure and its role in achieving Ethereum's stateless goal, an artifact currently confined only to implementation papers.

## 2 Layer 2 Solutions

To address the scalability limitations of Layer 1 (L1) blockchains, various off-chain solutions, collectively known as Layer 2[30] (L2), have been developed. These solutions aim to handle a large volume of transactions off the main chain while still leveraging its security.

### 2.1 Sidechains

A sidechain in decentralized scientific community is treated as an independent blockchain that functions parallel to the main chain. It has its own consensus mechanism and can use different rules than the main chain, offering a high degree of flexibility and scalability. For example, the Polygon network operates as a sidechain to Ethereum. One of the primary advantages of sidechains is their ability to execute transactions at a much faster rate with lower fees. However, their security is independent of the main chain, which can introduce vulnerabilities.

### 2.2 Rollups

Rollups, the dominant L2 scaling solution, are categorized by their underlying security mechanism: Optimistic Rollups (ORs) and Zero-Knowledge Rollups (ZKRs). While both achieve scalability by executing transactions off-chain, their fundamental security and finality guarantees are diametrically opposed, creating a critical trade-off for L2 design.

ORs operate under an "innocent until proven guilty" assumption: the Sequencer posts a new state root to L1, and it is assumed valid by default. The security of an OR relies on Fraud Proofs, a game-theoretic mechanism where anyone can challenge an invalid state transition during a fixed Challenge Period (typically 7 days). This economic incentive structure, enforced by slashing a malicious Sequencer's staked collateral, is the core security mechanism of ORs.

In contrast, ZKRs operate under a "guilty until proven innocent" model. They use Validity Proofs cryptographic, mathematically certain proofs to demonstrate that all off-chain transactions are valid before the new state root is accepted by the L1. The L1 contract instantly verifies this proof, not the original computation. This cryptographic certainty removes the need for a delay period. The differences are summarized below:

#### 2.2.1 Comparison of Rollup Architectures

| Feature | Optimistic Rollups (OR) | ZK-Rollups(ZKR) |
| --- | --- | --- |
| **Validation Method** | Assumed valid ("innocent until proven guilty"). | Proven valid by cryptography ("guilty until proven innocent"). |
| **Security Mechanism** | **Fraud Proofs** (Game Theory/Economic Incentive) | **Validity Proofs** (Mathematical/Cryptographic Guarantee) |
| **Withdrawal Time** | **Delayed** (e.g., approx 7 days) due to the challenge period. | **Instant** (as soon as the proof is verified on L1). |
| **L1 Execution** | Re-executes only the **disputed step** of a transaction during a challenge. | Verifies the **validity proof** for the entire batch. |
| **EVM Compatibility** | Typically **high** (easy to migrate existing L1 smart contracts). | Historically complex, but rapidly improving with **zkEVMs.** |

## 2.3 Plasma

Plasma uses a hierarchy of child chains that periodically commit a Merkle root of their state to the main chain. This structure reduces the amount of data stored on the L1 and allows for high-frequency transactions. A key security challenge for Plasma is the "mass exit" problem explained in section 5, where a malicious operator can hold funds hostage, forcing users to initiate an exit transaction to the main chain to retrieve their assets.

**2.4 State Channels**: State channels, such as the **Lightning Network** for Bitcoin and the **Raiden Network** for Ethereum, enable users to conduct multiple transactions off-chain. Only the opening and closing transactions are recorded on the main chain, while all intermediate transactions occur privately between participants. This provides a solution for micropayments, as it significantly reduces transaction fees. Security is maintained through a **Hash Time Lock Contract [HTLC]**, which ensures that participants can't cheat by broadcasting an old state of the channel.

## 3. Hash Time Lock Contract (HTLC) & Mass Exit Initiation
## 3.1 HTLC Definition and Cryptographic Construction

An HTLC is a conditional payment mechanism that uses a hashlock and a timelock to enforce a multi-hop payment. It works as follows:

1. A sender creates a **secret** (pre-image) and provides its **hash** to the receiver.
2. The sender locks the funds with a **hashlock**, requiring the receiver to reveal the pre-image to claim the funds.
3. A **timelock** is also set, giving the receiver a specific amount of time to claim the funds. If the receiver fails to claim the funds before the deadline, the payment is returned to the sender.

*Encryption Scheme:* $H(R)=P$

- H here is represented as a cryptographic hash function (like: SHA-256).
- R is the secret preimage.
- P is the public hash value.

HTLCs are crucial for enabling trustless cross-chain transactions and routing payments through a network of state channels. This process ensures that the payment is either completed or returned, eliminating the need for trust between parties and making it ideal for multi-hop payments in networks like the Lightning Network. Companies like **ACINQ** and **Lightning Labs** use HTLCs to facilitate off-chain payments. HTLC also aids as the technical backbone for many atomic swaps and trustless transactions.

## 3.2 Mass Exit Initiation

### 3.2.1 Block Withholding Attacks

**Mass exit initiation[60]** is a defense against **block withholding attack(BWA)**[28][29] & a major challenge for Plasma[62][63]. This occurs when a child chain's validators fail to publish new blocks. The chain is assumed to be byzantine, if a malicious operator withholds a block, users must initiate a mass exit from the plasma chain to retrieve their funds. This process can be slow- in the order of weeks and expensive. The operator's role is to verify:

- Plasma state a block ahead of BWA
- Contruct exit transaction to gather signatures from all users.
- Broadcast mass exit initiation transaction MEIT to the root chain.

**3.2.2 Mass Exit Initiation Transaction (MEIT)**

This MEIT includes a complete **bitmap** of the state being withdrawn, allowing observers to challenge any invalid claims. This type of attack, particularly possible on the Lightning Network, where an adversary forces many honest users to close their payment channels simultaneously by making a high volume of malicious transactions. This creates network congestion on the L1 chain, causing delays and forcing users to pay exorbitant fees to broadcast their channel-closing transactions before the timelock expires.

**4 Beginning of Layer 2 Security**
**4.1 Layer 2 Scaling solutions that are popular today**

While Layer-1 (L1) blockchains like the above run their own consensus to agree on transaction order and block validity, Layer-2 (L2) systems *inherit security* from an underlying L1 but use different mechanisms to order transactions and prove correctness.

- **Arbitrum**[64] (Optimistic Rollup)
  Posts transaction batches to Ethereum and assumes them valid unless challenged via a fraud-proof. Ethereum's PoS finality anchors security, but transaction ordering is controlled by an L2 sequencer (currently centralized).

- **Optimism**[65] (Optimistic Rollup)
  Similar to Arbitrum, with a challenge window for fraud-proofs. Security relies on Ethereum PoS for dispute resolution, but ordering and availability are dependent on the L2's sequencer and data-posting strategy.

- **zkSync**[66][67], **StarkNet**[68] (ZK-Rollups)
  Post validity proofs to Ethereum. Ethereum's PoS verifies these succinct proofs, ensuring state correctness. Sequencers order transactions off-chain but cannot create invalid state transitions without failing proof verification.

**EigenLayer[69] (Restaking Layer)**
Not strictly a rollup, but a mechanism for shared cryptographic security. This protocol, exemplified by EigenLayer, allows Ethereum Proof-of-Stake (PoS) validators to "restake" their already-staked ETH collateral to provide security to additional Actively Validated Services (AVSs), such as decentralized oracles, data availability layers, sidechains[61] or bridges. This mechanism creates pooled cryptoeconomic security, where the validator's capital is exposed to new slashing conditions imposed by the AVS. The security model ensures that the cost of corruption (CoC) for attacking multiple AVSs must remain prohibitively greater than the profit of corruption (PoC), as a misbehaving operator faces a cascading forfeiture of their staked position across both the AVS and the main Ethereum chain. The system operates as an open marketplace where validators selectively allocate their restaked capital based on the AVS's specific yield and penalty parameters.

## 4.2 Consensus in Layer-2

In L2s, consensus on transaction ordering is often centralized (single sequencer) or semi-decentralized, and disputes are resolved via the L1. This separation means L2 throughput can be much higher, but liveness and censorship resistance depend on sequencer design, while correctness depends on the integrity of proofs and availability of data.

**Consensus & Security Comparison: L1 vs L2**

| Blockchain / Protocol | Consensus Mechanism | Finality Time | Max Theoretical Throughput | Security Dependency | Key Trade-offs |
|---|---|---|---|---|---|
| Bitcoin | Proof of Work (PoW) | ~60 min (6 blocks) | ~7 TPS | Secured by global mining hashpower | Extremely secure but energy-intensive, low throughput |
| Ethereum | Proof of Stake (PoS, post-Merge) | ~15 min[42] | ~178.6 TPS | Secured by staked ETH | Energy efficient, scalable base for rollups |
| Algorand | Pure Proof of Stake (PPoS) | ~2.82sec[43] | ~25,000 TPS | Staked ALGO, random committee selection | Fast, strong finality, limited adoption |
| Cardano | Ouroboros PoS | ~2 minute[44] | ~18.02 TPS | Staked ADA, probabilistic leader election | Formal proofs of security, slower practical throughput |
| Polkadot | Nominated PoS (NPoS) with BABE + GRANDPA | ~30 sec | ~100,000 TPS across parachains | DOT stake, validator elections | Flexible, shared security, complexity in cross-chain validation |
| Near | PoS | ~0.6 sec | ~16,000 TPS | NEAR stake, validator elections | NightShade Sharding for mainstream adoption, high speed and low costs. |
| Polygon(L2) | PoS | ~5 sec | ~714.3 TPS | Staked Native Token POL for Lower fees | Fast Finality and high security through ZkEVM, trust assumption consider possible validator collusion. |
| Solana | PoS + Proof of History (PoH) | ~12.8 sec | ~65,000 TPS | Validator set, staked SOL | High throughput, vulnerable to outages and centralization. |
| Arbitrium(L2) | Optimistic Rollup (fraud proofs on Ethereum) | 7 days (challenge window) for full finality using Bounded Liquidity Delay[45] | ~40,000 TPS (off-chain) | Token ARB, Inherits Ethereum PoS; relies on sequencer & fraud-proof liveness | High throughput, but sequencer is centralized today. |
| Optimism(L2) | Optimistic Rollup (fraud proofs on Ethereum) | 7 days | ~714.3 TPS | Ethereum PoS | Similar to Arbitrum, relies on sequencer centralization. |
| zkSync/ Starknet(L2) | ZK-Rollup (validity proofs) | 2h | ~992 TPS (scales with proving efficiency) | Ethereum PoS + ZK proof soundness | Strong security, heavy cryptographic overhead. |
| Eigen Layer (middleware) | Restaking (extension of Ethereum PoS) | Inherits Ethereum finality | Varies (depends on service) | Ethereum PoS + additional slashing conditions. | Extends security to other protocols, introduces correlated slashing risk. |

## 4.3 Understanding Layer 2 through state-of-the-art cryptographic primitives
### 4.3.1 Zero-Knowledge Proofs (ZKP): Foundations

A zero-knowledge proof (ZKP) is a cryptographic method where one party, the prover, can convince another party, the verifier, that a statement is true without revealing any information beyond the statement's validity. ZKPs must satisfy three properties:

1. Completeness: An honest prover can always convince the verifier if the statement is true.
2. Soundness: We define soundness given a dishonest prover cannot convince to a given verifier that a chosen statement is indeed false.
3. Zero-Knowledge: The verifier learns nothing about the secret information beyond the truth of the statement.

In the scientific domain of blockchain, ZKPs are particularly useful for enhancing privacy and scalability, allowing for confidential transactions and the compression of large batches of off-chain transactions into a single, verifiable proof. Key steps to verify a zero knowledge proof include.[49]

1. **Proof generation** – a cryptographic ZKP is generated, showing that the new state (balances, contracts, etc.) is valid according to blockchain's rules.

2. **On-chain verification** – the ZKP is submitted to onchain L1, PoS validators only need to check the small proof, not re-run every transaction.

3. **Security guarantee** – if the proof verifies, it is mathematically impossible for invalid state transitions to be probabilistically possible.

### 4.3.2 Mathematical Computations behind ZKPs

The magic of ZKPs lies in a series of mathematical steps that transform a large, complex computation into a simple, verifiable proof. This process can be broken down into two main phases: **Proof Generation** and **Proof Verification**.

#### 4.3.2.1 Proof Generation (by the Prover)

The prover's task is to take a statement and its private "witness" (the secret information) and encode them into a mathematical object, typically a **polynomial**. This is the process of **arithmetization**. The prover takes the computation (often represented as an arithmetic circuit) and converts it into a series of polynomial equations. For the proof to be valid, all these equations must hold true.

For example, to prove knowledge of a secret number x such that $x^3 + 5 = 130$, the prover first creates a circuit, which is then converted into a polynomial identity, such as $P(x) = x^3 + 5 - 130$. The prover's goal is to prove they know a value of x for which this polynomial equals zero.

The prover then uses a **Polynomial Commitment Scheme (PCS)** to "commit" to this polynomial. A PCS is a cryptographic primitive that allows the prover to create a short, non-interactive hash-like representation of the polynomial. This commitment is **binding** (the prover can't change the polynomial later) and **hiding** (the verifier can't learn the polynomial's coefficients from the commitment alone). The most common PCS used in pairing-based ZKPs is [**KZG) (Kate-Zaverucha-Goldberg)**, which leverages elliptic curve pairings[SEC1][SEC2][32].

#### 4.3.2.2 Proof Verification (by the Verifier)

The verifier receives the proof, which includes the polynomial commitment and some additional information (the "witness"). The verifier does not need to re-run the original computation. Instead, they need to check a single, succinct pairing equation.

The core of the verification is to ensure that the polynomial P(x) from the prover's side indeed evaluates to zero at the correct point. Using the properties of pairings, the verifier can perform a single check of the form e(A, B) = e(C, D). This pairing check essentially verifies that the polynomial identity holds true at a randomly chosen point. The bilinear property of pairings is what makes this check possible.

A **pairing** is a function e(P, Q) that takes two points on elliptic curves (P from a group G1 and Q from a group G2) and maps them to a single element in a new group GT. It has the property that $e(aP, bQ) = e(P, Q)^{ab}$ (more details in the next section), which allows the verification process to check multiplicative relationships between values that are "hidden" in the elliptic curve points.

By combining the KZG commitment with pairings based functions, the verifier can be convinced that the prover knew the correct secret without ever learning what the secret was. The final proof is just a few elliptic curve points and field elements, making it extremely compact and easy to verify, regardless of the complexity of the original computation.

**SNARKS** were built on ZKP, which assumes an NP relation R, between given a witness w, and an instance x. i.e. $(x,w) \in R$. this relation is verifiable in polynomial time by a Turing machine. 3 stages of SNARK based ZKP are-

- Setup(pp) → (pk, vk) Setup Phase: Here input pp are public parameters, used for compute. The output generates proving and verification keys pk and vk, respectively.
- Prove(pk,x,w) → $\pi$. Proof Generation: Inputs required to generate proof, given the proving key pk, the witness w, and the instance x, using the assumption mentioned above, to output a proof $\pi$.
- Verify(vk,x,$\pi$) → 0/1. Verification: To verify the proof generated in step 2, given verification key vk, the instance x, and the proof $\pi$ as input, output is a binary- 1 if the proof is valid and 0 otherwise.

## 5. Mathematical Computation Model of ZKPs
### 5.1 Assumptions for ZKP Security
#### 5.1.1. Discrete Logarithm (DL) Assumption

The **Discrete Logarithm (DL) problem** is a fundamental assumption in modern cryptography. Given a base element g and a point h in a cyclic group, it's computationally infeasible to find the integer x such that $g^x = h$. This assumption underpins the security of many elliptic curve cryptosystems and is essential for the "hiding" property of polynomial commitment schemes. In simpler terms, it's easy to perform multiplication on the curve (x * G = X), but it's extremely difficult to reverse the process and find x from X and G.

#### 5.1.2. Computational Diffie-Hellman (CDH) Assumption

The **Computational Diffie-Hellman (CDH) assumption**[38] is a more specific variant of the DL problem. It posits that given a generator G and two points aG and bG (where a and b are unknown scalars), it is computationally infeasible to compute abG. This assumption is critical for the "binding" property of many commitment schemes. It ensures that a prover cannot forge a valid proof by manipulating the commitment.

### 5.2 Assumptions for Pairing-Based ZKPs

Pairing-based ZKPs, which are common for their efficiency, rely on an additional, stronger assumption related to bilinear pairings.

### 5.2.1 Bilinear Diffie-Hellman (BDH) Assumption

The **Bilinear Diffie-Hellman (BDH) assumption** is central to the security of pairing-based schemes like [KZG]. It states that given a generator P and the points aP, bP, and cP (where a, b, and c are unknown scalars), it is computationally infeasible to compute the value $e(P, P)^{abc}$. This assumption allows the verifier to check the relationship between multiple committed polynomials with a single pairing check. The "bilinearity" property of the pairing function $e(aP, bQ) = e(P, Q)^{ab}$ is a powerful tool for these proofs, and the BDH assumption ensures that this tool can't be exploited by a malicious prover.

## 5.3 Pairing-Friendly Curves

Zero-Knowledge Proofs, particularly the most efficient forms like **zk-SNARKs**[23] (Zero-Knowledge Succinct Non-Interactive Arguments of Knowledge), heavily rely on **pairing-friendly elliptic curves**[22]. These curves are special because they support a mathematical operation like **bilinear pairing**[36][37], allowing powerful cryptographic checks: it enables the verifier to check the relationship between multiple cryptographic commitments or keys in a single operation, which is crucial for the succinctness and efficiency of the proof.

### 5.3.1 Comparison of Pairing-Friendly Curves

| Curve | Key Properties | Improvement Over Previous Curve |
| --- | --- | --- |
| MNT Curves | **Type**: Early family of curves, with embedding degrees 3, 4, or 6. **Use**: Pioneering in pairing-based cryptography. | The MNT family was a foundational step in finding curves suitable for pairings, but they had limited security and flexibility. |
| BLS12-381 | **Type:** A Barreto-Lynn-Scott (BLS) curve with an embedding degree of 12. **Key Parameters:** A 381-bit prime field, and a 255-bit subgroup order. Use: Widely adopted in ZKPs (e.g., Zcash, Ethereum 2.0). | **Security & Efficiency:** BLS12-381 provides a more robust 128-bit security level, which was a significant improvement over the older BN254 curves that were found to have compromised security. Its low Hamming weight parameters optimize pairing computations, and its subgroup order is perfectly sized for 64-bit machine arithmetic, making it highly performant for zk-SNARKs. |
| BLS12-461 | **Type:** Another BLS curve with an embedding degree of 12. **Key Parameters:** A 461-bit prime field. **Use:** Proposed as a more secure alternative to BLS12-381. | **Increased Security Margin:** BLS12-461 provides a higher security margin, targeting a more conservative 128-bit security level against advanced cryptanalytic attacks that may threaten curves with smaller field sizes. |
| BN-462 | **Type**: A Barreto-Naehrig (BN) curve. **Key Parameters:** A 462-bit prime field. Use: A competitor to the BLS12-461. | **Alternative Construction:** BN curves offer a different approach to constructing pairing-friendly curves. BN-462 is a competitor to the BLS12-461 with equivalent security. The choice between them often comes down to specific implementation details and performance characteristics for a given application. |
| BLS48-286 | **Type:** A BLS curve with a high embedding degree of 48. **Key Parameters:** A 286-bit prime field. Use: Research and advanced protocols. | **Improved Prover Time**: By using a higher embedding degree, BLS48-286 can sometimes offer better performance for the prover, as some cryptographic operations become more efficient at the cost of a larger prime field. This makes it a potential candidate for future proof systems. |

The curve currently most widely used in ZKP computations is **BLS12-381**[21][39]. It is employed in the final **proof verification step**. During this step, the verifier receives a succinct proof from the prover and performs a pairing check. The pairing operation verifies the integrity of the complex polynomial equations that represent the computation performed by the prover. It essentially allows the verifier to confirm that the prover correctly transformed the input data (the "witness") into the final proof without revealing the witness itself. This single, efficient pairing check is what makes zk-SNARKs so powerful and fast to verify, regardless of the complexity of the original computation.

While the Fotiadis–Martindale[17], Freeman–Scott–Teske[18][19], and Kachisa–Schaefer–Scott (KSS)[20] curves are also 128-bit security level pairing-friendly curves[16], they can be used for cryptographic applications, but they are not as widely adopted as BLS12-381 due to a variety of reasons. The primary reason is that BLS12-381 has become a de facto standard in the ZKP and blockchain space, particularly in protocols like Zcash[26], Ethereum, and Filecoin. This widespread adoption is due to its optimal balance of security, efficiency, and implementation properties.

Other curves, while theoretically sound, often have characteristics that make them less practical for broad use. For example, some may offer slightly better security but are much less efficient to compute. Others might be harder to implement or lack the "low Hamming weight" property of the BLS12-381 parameter, which is a key factor in speeding up the Miller loop, a critical step in pairing computations. The cryptographic community, particularly in the blockchain space, has largely converged on a few well-tested and highly-optimized curves to ensure interoperability and security.

**6 Security Primitives of ZKPs in Layer 2**

ZKPs guarantee L2 security by providing **mathematical certainty** that only valid state updates are committed to Ethereum, even if the L2 operator is untrusted. This gives ZK-Rollups stronger security assurances and faster finality compared to Optimistic Rollups. Some Layer-2 solutions relying on ZKPs are **zkSync** or **StarkNet** use **ZK-Rollups**.

- **No trust in sequencers:** Even if the L2 sequencer is malicious, they cannot submit invalid state — the proof would fail.

- **Finality speed:** Once Ethereum accepts the proof, the L2 transactions are as final as an L1 transaction.

- **Data integrity:** Users can reconstruct state from published data + proofs, ensuring no hidden manipulation.

Understanding the technical details of ZKP algorithms, such as the trade-offs between different cryptographic curves and the impact of hardware optimizations, is crucial for grasping the promises of ZK-Rollups. These promise untrustworthy sequencers, immediate finality, and provable data integrity are not just theoretical beliefs; they are practical assurances built on the mathematical certainty provided by these underlying cryptographic primitives. Therefore, diving into benchmarks like R1CS and comparing various ZKP frameworks is essential. It provides concrete evidence of how different implementations deliver on the promised speed and security, allowing developers and users to critically evaluate which solutions truly live up to the Layer 2 vision.

**6.1 R1CS as a Foundation for ZKPs**

Rank-1 Constraint Systems (R1CS) remain one of the most widely used intermediate representations for zero-knowledge proofs, since they allow general computations to be expressed as algebraic constraints. Early benchmark studies (e.g., Spartan 2020[SPA]) showed how different proof systems scale with growing R1CS instance sizes, highlighting the trade-offs between interactive and non-interactive proof styles. Interactive proofs often incurred higher verification costs but maintained smaller proofs, while non-interactive SNARKs and NIZKs demonstrated succinct proofs with lower verifier overhead. Although these results are not the latest, they illustrate that R1CS complexity fundamentally shapes proving time, proof size, and verification efficiency—trends that continue to influence today's proof systems and framework design.

## 6.2 Understanding Plonks improvements over Sonic: how Groth16 redefined universal SRS.

The trusted setup stage in a zkp generates a public, setup-dependent parameter required by pairing-friendly curves[Gro16] called a Structured Reference String(SRS). A term now preferred over Common Reference String (CRS). Historically, most zk-SNARKs required a **trusted setup** that produced an SRS specific to a single circuit or relation. This meant any change to the circuit required a new setup. The first key breakthrough was Groth et al.'s[GKM+] introduction of a **universal and updatable** SRS, allowing a single setup to support all circuits up to a bounded size, but this string scaled *quadratically* in size, making it impractical for large circuits.

[Sonic] provided the first potentially practical zk-SNARK with a **linear-size**, universal, and continually updatable SRS, solving the quadratic-scaling issue of prior universal setups. However, the fully-succinct version of Sonic still had relatively high proof construction overheads.

[PLONK] (Permutations over Lagrange-bases for Oecumenical Noninteractive arguments of Knowledge) improvised by introducing a universal SNARK with **significantly lower prover running time** (roughly 7.5 to 20 times fewer group exponentiations than fully-succinct Sonic). Plonk is a modern zk-SNARK protocol that uses a fundamentally different arithmetization.

## 6.3 "Plonkish" arithmetization

PLONK introduced the "Plonkish" arithmetization, replacing R1CS with a system based on **Custom Gates** and a **Permutation Argument**.

- **Custom Gates:** A single gate equation can be more complex than R1CS. This flexibility allows defining operations like cryptographic hashes or bitwise logic in one constraint, making circuits **smaller** and **proving faster** for certain operations (e.g., Pedersen Hash).
- **Permutation Argument:** It uses a check based on the **Grand Product Argument** (an evolution of Sonic's technique) to ensure consistency of wires connecting different gates.

  Example of a **Custom Gate Equation** in PLONK arithmetization- $Q_L \cdot a + Q_R \cdot b + Q_M \cdot a \cdot b + Q_O \cdot c + Q_C = 0$). This single equation defines the allowed operation for one row (or gate) in the circuit's execution trace.

| Variable | Type | Definition |
| --- | --- | --- |
| a,b,c | Witness values | These represent the **wire values** or **signals** (inputs and outputs) for a single gate in the circuit's execution trace. They are the values the Prover is trying to prove knowledge of. |
| $Q_L, Q_R, Q_M, Q_O, Q_C$ | Selector Polynomials | These are fixed, **public constants** chosen by the circuit designer (the compiler/trusted setup) for each gate. They are often referred to collectively as Q. |
| $Q_L, Q_R$ | Left/Right Selectors | They select the variables a and b for the linear terms. |
| $Q_M$ | Multiplication Selector | It selects the term $a \cdot b$ to enable a multiplication operation. |
| $Q_O$ | Output Selector | It selects the output variable c. (Note: In some definitions, c is the output of the gate and QO is sometimes QOUT or similar, and is often assigned a negative value to move it to the right side of the equation). |
| $Q_C$ | Constant Selector | It allows a **constant** to be added to the equation. |

## 6.4 How the Plonk Equation Works

The equation is structured to be a generalized way to encode any quadratic (or degree-2) constraint, where all terms must sum to zero:

- **Linear Terms (Addition):** $Q_L \cdot a + Q_R \cdot b$
- **Quadratic Term (Multiplication):** $Q_M \cdot a \cdot b$
- **Output Term:** $Q_O \cdot c$
- **Constant Term:** $Q_C$

By setting the selector constants (Q) appropriately for a specific gate, the circuit can perform any operation:

1. **Multiplication Gate (a·b=c):**
   - Set $Q_M=1$, $Q_O=-1$, and all others to 0.
   - Equation becomes: $(1 \cdot a \cdot b) + (-1 \cdot c) = 0 \Rightarrow a \cdot b = c$.
2. **Addition Gate (a+b=c):**
   - Set $Q_L=1$, $Q_R=1$, $Q_O=-1$, and $Q_M=Q_C=0$.
   - Equation becomes: $(1 \cdot a) + (1 \cdot b) + (-1 \cdot c) = 0 \Rightarrow a + b = c$.
3. **Custom Gates:** By assigning more complex values to the Q polynomials, PLONK can enforce higher-degree constraints or specialized operations (like checking if a variable is a boolean $a \in \{0,1\}$).

The **PLONK Prover** must find values for a,b,c (the witness) such that for all rows in the circuit, this equation with row-specific public Q values is satisfied.

**Note:** The variable r is not part of this gate equation but is typically used as a **random challenge** value sent by the Verifier to the Prover in the interactive stages of the protocol. It is crucial for ensuring the soundness of checks like the Permutation Argument.

Plonks achieved simplifying the algebraic constraints, moving the focus to **"Evaluations on a subgroup rather than coefficients of monomials,"** which streamlined both the permutation argument and the circuit arithmetization step.

## 7 Computational analysis of various ZKP frameworks

[zkBench] Includes 13 different elliptic curves[34][35] implemented across 9 libraries to perform arithmetic operations in finite fields benchmarks across 5 ZKP frameworks. The framework allows adding circuit implementation with arbitrary payloads in the backend to measure execution time and performance benchmarks based on 2 log parsers for visualization and information extraction- [zkalc] & [zk-Harness].

From the paper we learn, field-arithmetic performance is **library–curve dependent** rather than monotonic with field size. With BN254, **gnark-crypto**—leveraging Botrel et al.'s[BOT] Montgomery-multiplication optimizations and Longa's[LON] inner-product—delivers about **30% lower latency** for both addition and multiplication than other implementations. On **BLS12-381**, however, gnark-crypto is on average **~9% slower** than competing libraries, so BN254's smaller modulus does **not** guarantee universal speedups. Across all stacks, **addition ≪ multiplication** in cost, but the **winner flips by curve**, underscoring that practitioners must choose the **curve+library pair** deliberately.

Historically, curves with smaller field characteristics like **BN254** were initially **believed to be inherently faster** than larger-field curves like BLS12-381 for arithmetic operations

## 7.1 Comparison of ZKP Frameworks and Arithmetic Libraries

| Framework / Library | Arithmetic Backend | Circuit Efficiency | Proof Size | Speed Optimizations | Hardware Sensitivity |
|---|---|---|---|---|---|
| **Groth16 (with gnark-crypto)** | BN254, BLS12-381 | Very efficient with optimized libraries | Smallest proofs | Fast for scalar arithmetic, MSM up to 1.83× faster than blstrs | Gains up to 50% on compute-optimized hardware |
| **PlonK** | BLS12-381 | Flexible circuits, some overhead | Small–medium | Optimized circuits reduce proving time | Moderate sensitivity |
| **Halo2** | BLS12-381 | Dependent on circuit design (can be slower or faster) | Medium | Circuit-specific optimizations make it faster in some cases (1.09× faster in second test) | Moderate |
| **STARKs (starky)** | Prime fields, FFT-friendly-Goldilock fields | High scalability | Very large (299×–679× bigger than Groth16/PlonK) | Parallelism benefits proving time | Strong dependence on memory bandwidth (40% gains possible) |
| **snarkjs (JS-based)** | ffjavascript | Lower efficiency | Small–medium | 14× slower than gnark in SHA256 proving | More limited optimization potential |
| **gnark-crypto** | BN254, BLS12-381 | Very efficient MSM operations | Small | 9.4×–13× faster vs ffjavascript in G1/G2 | Significant benefit from compute-optimized hardware |

## 8 Performance Improvement in commitment schemes curated by Ethereum
## 8.1 Merkle Sum Tree

Built around the problem that public parameters should be easy to produce and verify. Binance adopted vitalik's idea of Merkle Sum Tree[SMST]- proof of liability. Merkle tree is a characteristic datastructure where inner nodes represent balanced sum information from the child tree. This demonstrasted critical privacy leakages for users and created possibilities of taking away user anonymity, which is a core fundamental pillar of blockchain security.

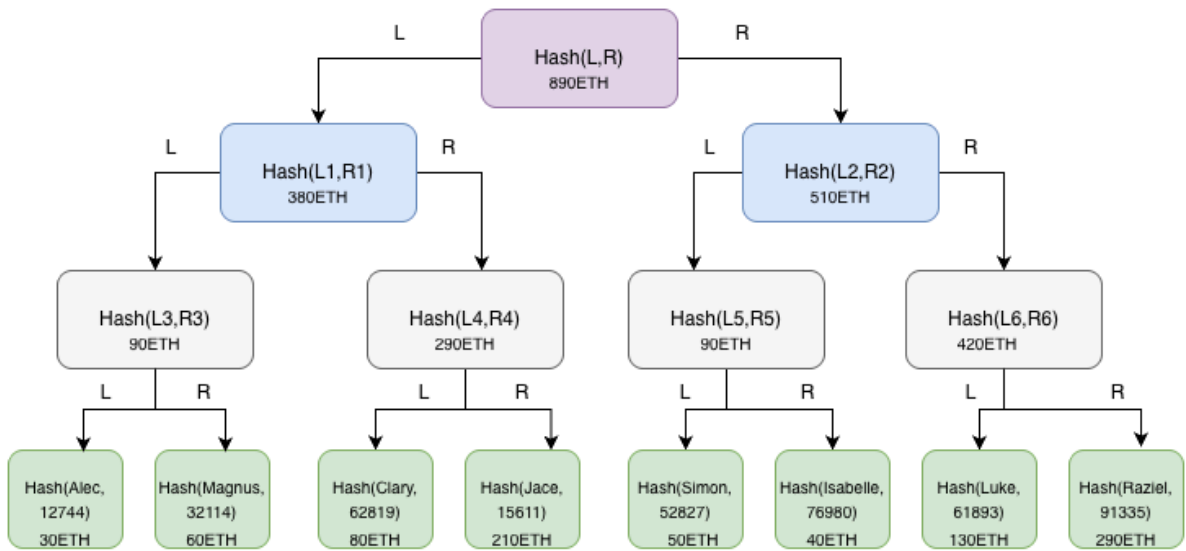

The diagram illustrates the structure of an Ethereum-inspired Merkle Sum Tree (MST). At the lowest level, each leaf node corresponds to an individual user's record and contains:

1. A pseudonymous user label,

2. A random salt to prevent preimage attacks, and

3. The exact balance held by that user.

Each leaf is hashed, and pairs of leaves are combined into internal nodes. Every internal node stores two pieces of information:

- a cryptographic hash of its left and right children, and

- the sum of the balances in its subtree.

As this process continues upward, the tree produces a single root hash along with a cumulative total liability value. This enables efficient verification: a user can check that their leaf is included in the root commitment by verifying only the nodes along their path, rather than the entire dataset.

**8.1.1 Privacy Risks: Mitigated by STARKS**

However, the approach introduces measurable privacy risks. Because each internal node reveals the aggregated balance of all users in its subtree, adversaries can infer partial information about user clusters, especially when the user distribution is sparse or unbalanced. This reduces anonymity and contradicts one of the foundational security assumptions of blockchain-based systems: minimizing unnecessary leakage of user state.

SNARK ZKPs adopted vector based KZG-IOP commitment schemes which were much safer and mitigated the security concerns but extended problems of it's own with large proof sizes. The largest trusted setup ever built **perpetual powers of [tau] ceremony** with the effective size of $2^{28} \approx 270$ million highlights scalability challenges with the expected proof volume size requirements of layer 2 users. KZP-IOP has a slow prover overhead due to multi-scaler multiplication performed for $2^{28}$ points with polynomial arithmetic operations(fast fourier transform) involved in encode, interpolate and polynomial evaluation assumed in large finite fields(like BN254 or BLS13-281). There are additional concerns- larger the trusted setup, larger the computation waste, larger the trust assumption making it a huge effort to achieve on a massive scale.

Poseidon hashes over goldilock field bypass the need for a large trusted setup, with a more efficient small commitment size and faster verification time. This is because its parameters are deterministically defined and independent of any secret randomness. Unlike pairing-based SNARKs that rely on trapdoor-based structured reference strings, Poseidon's field arithmetic and transparent design allow it to integrate into proof systems (STARKs, FRI-based IOPs[41]) that are trustless by construction.

Empirical extrapolation suggests that executing a $2^{28}$-sized Poseidon hash computation over the Goldilocks field requires approximately **322 seconds** on an Apple M1 processor. Considering the architectural advancements of the Apple M4—namely, higher clock frequency, wider vector pipelines, and improved memory bandwidth—a **2.5× performance gain** can be reasonably anticipated. The expected computation time can thus be approximated as:
$$T_{M4} = T_{M1} / \text{Speedup} = 322/2.5 \approx 129 \text{ seconds.}$$

Accordingly, the runtime for the same workload on the Apple M4 is projected to fall within **120–150 seconds**, reflecting a **2–3× improvement** depending on memory-bound or compute-bound workload characteristics. This estimation provides a baseline for assessing scalability of Poseidon-based hashing within large-field cryptographic computations.

## 8.2 Verkle trees for ethereum's stateless future
### 8.2.1 Merkle Particia Trie

At the time of writing this paper consensus nodes consume the entire state of the blockchain to attest on proposed blocks, this state burden has extended the high harden requirements for ethereum community, making access to large and fast memory a difficult problem to solve. Merkle Patricia Trie stores key-value data, has about 9 levels of depth comprising of 3 types of nodes, **root extension nodes**- indexing common prefix keys from the account addresses, **branch nodes** representing a hexadecimal distribution from 0 to f and **leaf nodes** indexing final remaining nimble from the hash of a typical ethereum account address. Each node in a Merkle Patricia Trie is indexed by the hash of its contents. To prove a leaf belongs to the trie, the prover must recreate it's path from the root. Ethereum maintains 4 separate Mekle Patricia Tries each with a separate use for each node on the blockchain.

- State root: Trie is indexed by the address of account, stores balance with nonce; For smart contracts, it also stores code hash and storage root tracking smart contract storage variables.
- Transaction root: Containing transaction data.
- Receipts root: Containing transaction receipt data.
- Withdrawals root: Containing proof of state withdrawals.

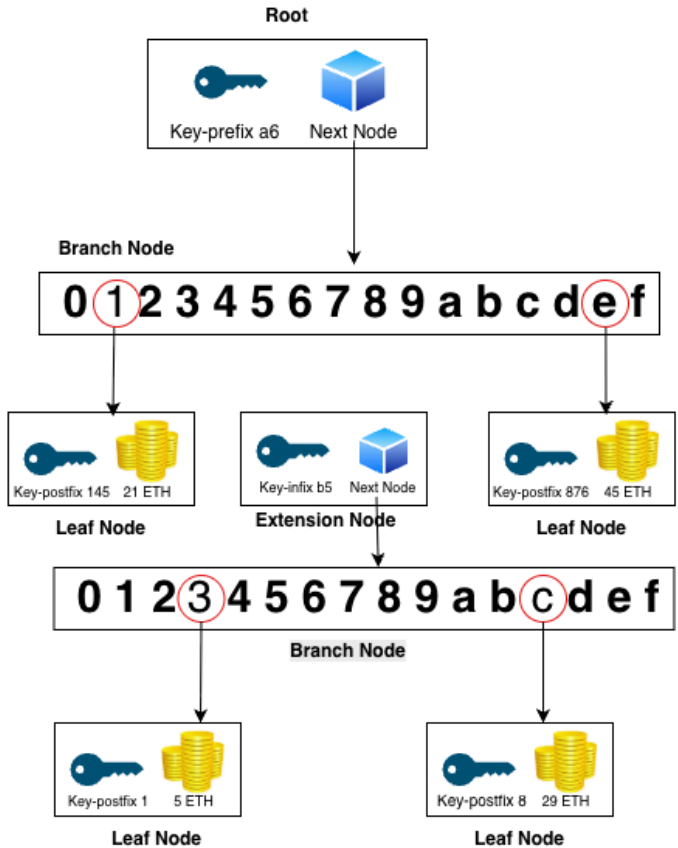

| Key | Value |
|---|---|
| a61145 | 21 ETH |
| a6e876 | 45 ETH |
| a6b5c8 | 29 ETH |
| a6b531 | 5 ETH |

### 8.2.2 Verification Algorithm of Merkle Proofs
Algorithm for verifying merkle proofs- The algorithm runs in O(log N) time where N is the number of leaves.

Algorithm 1: VerifyMerkleProof(leaf, proof[], root)
Input:
  leaf — data element to verify
  proof — list of sibling hashes along the path to the root
  root — expected Merkle root hash

1: h ← Hash(leaf)
2: for each sibling s in proof do
3:    if s.position == "left" then
4:       h ← Hash(Concat(s, h))
5:    else
6:       h ← Hash(Concat(h, s))
7:    end if
8: end for
9: if h == root then
10:    return True  // proof is valid
11: else

12:     return False  // proof is invalid
13: end if

Proof Size for a single account: Given Trie depth is 9 levels on average, branch nodes are 16 nimbles [0 -> 9, a -> f] and keccak256 hash function has a constant size of 32 bytes.
**Proof Size = Depth x nimble x proof size = 9 X 16 X 32** ≈ 4 KB

For 1000 accounts this is approximately 4MB**,** this massive data problem is what ethereum is trying to solve using **Verkle Trees**.

### 8.2.3 Verkle Trees: How does ethereum makes a more efficient data structure with vector commitments?
In the earlier sections we explained how KZG commitments schemes are used as vector commitments.
Given data vector $[e_1, e_2, e_3, ....e_i]$, The prover sends a commitment C of vector $(e_1, e_2, e_3, ....e_i)$ to a verifier without revealing it and to prove the vector contains data $e_i$ at index i, he also shares a generated proof($e_i$) of the choosen index with witness $w(e_i)$. This convinces the verifier of the prover's claim without ever revealing the entire vector data.

Verkle trees need a vector commitment scheme, some existing implementations[Verkle] utilize the KZG polynomial commitment schemes using rust-verkle implementing BLS 12381 elliptic curve defined over a large prime field ≈ $2^{381}$ bits in length, nearly 48 bytes in size. The committed vector elements belong to a scalar field ≈ $2^{255}$ bits nearly 32 bytes in size. The curve further supports 2 subgroups($G_1, G_2$) needed to define pairings for KZG verification.
$$e(G_1, G_2) \rightarrow G_T$$
Ethereum [Verkle] Tree doesn't use KZG commitment schemes due to restrictions over trusted setup and to avoid pairing based verifications. Ethereum uses **Inner Product Argument commitment(IPA)**[72][73] introduced by Bootle et al. [BCCGP16] also implemented in Bulletproofs [BBBPWM18], IPA uses a BanderSnatch curve defined over BLS 12381 in a **scalar** field making the base field much smaller and under 32 bytes as compared to KZG. Vector value commitment C belong to the scalar field ≈ 253 bits, also under 32 bytes. **IPA commitment** to the vector is a single elliptic curve point like KZG. **IPA proofs** however contain multiple elliptic curve points, so with constant commitment size the proof size grow logarithmically with the vector size, $2\log_2(n)$ each under 32 bytes.

### 8.3 Table: Comparison of KZG and IPA Vector Commitment Schemes

| Commitment Scheme | Curve & Field | Vector Element Field Size | Commitment Type | Proof Structure | Proof Size | Trusted Setup |
|---|---|---|---|---|---|---|
| **KZG Commitment** | BLS12-381 (Pairing-based) | Scalar field ≈ $2^{255}$ bits (~32 bytes) | Single elliptic curve point(2381bits) < 48 Bytes | Single elliptic curve point | < 48 Bytes, Single EC point | **Required** (SRS) |
| **IPA Commitment** | BanderSnatch (Non-pairing) | Scalar field ≈ $2^{253}$ bits (<32 bytes) | Single elliptic curve point(2255 bits) < 32 Bytes | Inner Product Argument (multiple elliptic curve points) | each < 32 bytes, $2\log_2(n)$ EC points | **Not required** (Transparent) |

## 9 Limitations and Future work

While this survey consolidates existing understanding of Layer-2 security, several frontiers remain open for detailed exploration. The next version of this work will extend the discussion on Verkle trees, the emerging vector-commitment structure envisioned for Ethereum's stateless architecture. Specifically, future work will decompose the internal organization of Verkle trees, analyze spatial complexity, and quantify verifier and prover performance under both KZG-based and Inner-Product Argument (IPA)-based implementations.

We plan to formalize the comparative impact of these designs on state storage efficiency, proof generation latency, and gas economics within Ethereum's roadmap. Furthermore, subsequent iterations of this survey will expand the quantitative analysis of cryptographic performance trade-offs in commitment schemes, providing empirical benchmarks and implementation insights. Through these extensions, the follow-up paper will aim to serve as a definitive reference for the security economics and cryptographic scalability primitives shaping next-generation Layer-2 and stateless blockchain systems.